\documentstyle[12pt]{article}
\begin{document}
    rightline{TUHE-9662, HEP-PH/9606407}
\begin{center}
\mbox{{\Large\bf The Proper Identification of the Heavy Baryons}}
\vspace{.1in}\\
Jerrold Franklin\\
{\it Department of Physics,Temple University,\\
Philadelphia, Pennsylvania 19122}\\
June, 1996
\end{center}
\begin{abstract}

The new designation of heavy baryons proposed by Falk is shown to violate
minimal symmetry sum rules.
\end{abstract}
\vspace{.2in}

A new interpretation has been proposed by Falk\cite{falk} for the heavy baryons
because
the conventional assignments violate mass relations that Falk bases on heavy
quark effective theory and SU(3) symmetry.   However,
we show here that Falk's proposed new baryon designations violate more general
mass sum rules\cite{cb,acb} that are based on a
smaller set of assumptions than those used by Falk.  The more general sum rules
are satisfied by the conventional baryon
assignments.

Three of these sum rules that can be tested with the present set of observed
heavy baryons are
\begin{equation}
(\Sigma^{*0}-\Lambda^0)+\frac{1}{2}(\Sigma^0 -\Lambda^0)=
(\Sigma_c^{*+}-\Lambda_c^+)+\frac{1}{2}(\Sigma_c^+ -\Lambda_c^+)
\end{equation}
\hspace{1.3in}$(307_{-10}^{+23})$\hspace{1.4in}$(330\pm 7)$\\
\begin{equation}
\hspace{1.75in}
=(\Sigma_b^{*0}-\Lambda_b^0)+\frac{1}{2}(\Sigma_b^0 -\Lambda_b^0).
\end{equation}
\hspace{3.3in}$(316\pm10)$\\
\begin{equation}
\Sigma^+ +\Omega^{-} -\Xi^0 -\Xi^{*0} =
\Sigma_c^{++}+\Omega_c^{0} -2\Xi_c^{\prime +}.
\end{equation}
\hspace{1.5in}$(15_{-18}^{+12})$\hspace{1.3in}$(27\pm 30)$\\
\vskip .1in
\noindent	
The experimental value in MeV of each baryon sum is given below each equation,
using the heavy baryon masses given in
Ref.\cite{falk} with the conventional assignments.  The $\pm$ values for the
light baryons result from the fact that three different
light baryon combinations could be used in each equation.\cite{acb}

These sum rules require no symmetry assumptions about interactions or wave
functions.  They are based on two assumptions:\\
\begin{enumerate}
\item The ground state baryons are composed of three spin one-half colored
quarks whose spins add up to the baryon spin with no
orbital angular momentum.\\
\item The interaction energy of each pair of quarks does not depend on the third
quark in the baryon.
\end{enumerate}
Both of these assumptions are used (at least, implicitly) in deriving Eqs.
(1)-(7) of Ref.\cite{falk}.
A number of additional assumptions are also made in deriving some of the seven
equations in Ref.\cite{falk}.  Note, however, that
the equality between the c- and b-baryon combinations in our Eqs. (1) and (2) is
the same as Eq. (2) in Ref.\cite{falk}.  That
relation does not depend on any heavy baryon assumption, and we see no reason to
separate it from the light baryon sector.

Based on the failure of two of the relations in Ref.\cite{falk}, Falk has
proposed alternate assignments for the $\Sigma_c^{*}$,
$\Sigma_c$,  $\Sigma_b^{*}$, and $\Sigma_b$ baryons.  He proposes that the heavy
baryons that have been considered as the $\Sigma$ baryons, in fact are the
$\Sigma^{*}$ baryons. The actual heavy $\Sigma$ baryons
would then be lower in mass, and not as yet detected.  However if the
$\Sigma_c^{*}(2530)$ baryon in Eq. (1) above is replaced by
the $\Sigma_c(2453)$,
and the $\Sigma_c$ by a lower mass baryon,
then the right hand side decreases by at least 77 MeV to become $<$ 253 MeV.
Our sum rule (1) is then no longer in reasonable agreement.
Putting in Falk's proposed $\Sigma_c(2380)$ further lowers the right hand side
of Eq. (1) to 216 MeV, worsening the disagreement.
Replacing the $\Sigma_b^*(5852)$ by $\Sigma_b(5796)$ would change the right hand
side of Eq. (2) to $<$260 MeV, and using a new
$\Sigma_b(5760)$ would make it 224 MeV, so that the b-baryon sum rule would no
longer agree.
Using the proposed $\Sigma_c(2380)$ in Eq. (3) changes the right hand side to
-46$\pm${30} also putting it into disagreement.

We see that Falk's proposed heavy baryon assignments violate all three of the
sum rules above, while the sum rules are satisfied by
the conventional heavy baryons.

The two relations  that do not agree with the standard heavy baryon designations
are equations (3) and (7) of Ref.\cite{falk}, which
are
\begin{equation}
\frac{\Sigma_b^{*}-\Sigma_b}{\Sigma_c^{*}-\Sigma_c}
\hspace{.2in}=\hspace{.2in}
\frac{B^*-B}{D^*-D}
\end{equation}
\hspace{1.65in}$(0.73\pm.16)$\hspace{.65in}(0.33)\\
\vskip .1in
and
\begin{equation}
(\Sigma_c^{*+}-\Lambda_c^+)+\frac{1}{2}(\Sigma_c^+ -\Lambda_c^+)=
(\Xi_c^{*+}-\Xi_c^+)+\frac{1}{2}(\Xi_c^{\prime +}-\Xi_c^+).
\end{equation}
\hspace{1.15in}$(330\pm7)$\hspace{1.4in}$(223\pm9)$\\
\vskip .1in
\noindent
Equation (4) relies on assumptions (1) and (2) above, and also assumes that the
spin-spin interaction between pairs of quarks is
inversely proportional to the product of the quark masses, but otherwise
independent of quark flavor and baryon location.  The right
hand side of Eq. (4) is equal to $m_c/m_b$, the ratio of c-quark to b-quark
mass, as measured by the heavy meson mass combinations.

Equation (4) can be extended to the light baryons because it does not use any
assumption specific to heavy baryons.  That is, we can
write
\begin{equation}
(\Delta^+-p):(\Sigma^{*+}-\Sigma^{+}):(\Sigma_c^{*++}-\Sigma_c^{++})
:(\Sigma_b^{*+}-\Sigma_b^{+})
=\frac{1}{d}\hspace{.05in}:
\hspace{.05in}\frac{1}{s}\hspace{.05in}:
\hspace{.05in}\frac{1}{c}\hspace{.05in}:\hspace{.07in}\frac{1}{b}
\end{equation}
\hspace{.17in}$(297)$\hspace{.17in}:\hspace{.38in}$(194)$\hspace{.15in}:
\hspace{.15in}$(77\pm7)$\hspace{.2in}:\hspace{.2in}$(56\pm11)$
=(297):(192)(61):(20)\\
\vskip .1in
\noindent
where we have used the quark symbol for its mass. The numbers under the right
hand side of Eq. (6) have been normalized to the
$\Delta^+$-p mass difference in MeV, using the quark masses
\begin{equation}
u=d=330 MeV,\hspace{.1in} s=510 MeV,\hspace{.1in} c=1.6 GeV,\hspace{.1in} b=4.8
GeV.
\end{equation}

Looking at Eq. (6) we see that the $\Sigma_c^*-\Sigma_c$ mass difference is not
unreasonable with the conventional assignments, but
the $\Sigma_b^*-\Sigma_b$ difference is too large.  However there are a number
of three body effects that can be expected to modify
Eq. (6).\cite{dl}  In fact, it is generally true that relations involving spin
$\frac{3}{2}$-spin
$\frac{1}{2}$ mass differences are only accurate to about 20 MeV.  This is seen
in light baryon mass splittings\cite{acb,jf}, and in
heavy baryon mass splittings, as in Ref.\cite{acb} and Eqs.(2), (4), and (6) of
Ref.\cite{falk}  Taking this into account makes the
b-baryon mass difference look not quite as serious a problem as when a small
number is put into a ratio, as in
Eq.(4), magnifying the effect of the 36$\pm$11 MeV difference between the two
sides of Eq. (6) for $\Sigma_b^*-\Sigma_b$.

Falk describes Eqs. (4)-(7) of Ref.\cite{falk} as including corrections linear
in the strange quark mass.  However, in
equation (5) [equation (7) of Ref.[1]], which agrees badly with the conventional
heavy baryon assignments, the mass splitting of the left
hand side is proportional to 1/ud, while the right side mass splitting is
proportional to 1/ds.  This results in a
relatively large difference in the two sides of Eq. (5), proportional to the
quark mass ratio (s-u)/uds
which persists even in the heavy quark limit.  In
fact, since the baryon combinations in Eq. (5) were chosen to cancel out the
heavy quark spin contribution, they do not rely on any
heavy quark assumptions and should apply equally well (or badly) to the light
baryon sector.  If exactly the same assumptions that went into Eq. (5)
above were applied to the light baryons, then the relation
\begin{equation}
(\Sigma^{*0}-\Lambda)+\frac{1}{2}(\Sigma^{0}-\Lambda)=
\Xi^{*0}-\Xi^{0}
\end{equation}
\hspace{1.7in}$(307)$\hspace{1.5in}$(217)$\\
\vskip .1in
\noindent
could be written.  Equation (8) has just the same disagreement as Eq. (5).  We
conclude that the spin-spin correction term
proportional to
(s-u)/usd cannot be safely ignored in either the light or heavy baryon
sector.

Our conclusion is that there is no good reason to change the conventional
designation of heavy baryons, and strong evidence in Eqs.
(1)-(3)
against the heavy baryon assignents proposed in Ref. 1.

\end{document}